\documentclass[prl,twocolumn,amsmath,amssymb,superscriptaddress,showpacs]{revtex4-1}
\usepackage{graphicx}
\usepackage{color}

\newcommand{\beq}{\begin{equation}}
\newcommand{\eeq}{\end{equation}}
\newcommand{\be}{\begin{equation}}
\newcommand{\ee}{\end{equation}}
\newcommand{\beqn}{\begin{eqnarray}}
\newcommand{\eeqn}{\end{eqnarray}}
\newcommand{\bea}{\begin{eqnarray}}
\newcommand{\eea}{\end{eqnarray}}
\newcommand{\bearr}{\begin{array}}
\newcommand{\enarr}{\end{array}}

\newcommand{\comment}[1]{}

\begin{document} 

\title{Effective grand-canonical description of condensation in negative-temperature regimes}
      
\author{Stefano Iubini}
\email{ stefano.iubini@cnr.it}
\affiliation{Istituto dei Sistemi Complessi, Consiglio Nazionale
delle Ricerche, via Madonna del Piano 10, I-50019 Sesto Fiorentino, Italy}
\affiliation{Istituto Nazionale di Fisica Nucleare, Sezione di Firenze, 
via G. Sansone 1 I-50019, Sesto Fiorentino, Italy}

\author{Antonio Politi}
\email{a.politi@abdn.ac.uk}
\affiliation{Institute for Complex Systems and Mathematical Biology  
           University of Aberdeen, Aberdeen AB24 3UE, United Kingdom}
\affiliation{Istituto dei Sistemi Complessi, Consiglio Nazionale
delle Ricerche, via Madonna del Piano 10, I-50019 Sesto Fiorentino, Italy}

\begin{abstract}
The observation of negative-temperature states in the localized phase of the Discrete Nonlinear Schr\"odinger (DNLS) equation
has challenged statistical mechanics for a long time. 
For isolated systems, they can emerge  as stationary extended states  through a large-deviation mechanism occurring for finite sizes, while they are formally unstable in grand-canonical setups, being associated to an unlimited growth of the condensed fraction.
Here, we show that negative-temperature states in open setups
are metastable and their lifetime $\tau$ is exponentially long with the temperature,
$\tau \approx \exp(\lambda |T|)$ (for $T<0$).
A general expression for $\lambda$ is obtained in the case of  a simplified stochastic model of non-interacting particles.
In the DNLS model, the presence of an adiabatic invariant,  makes $\lambda$ 
even larger because of the resulting freezing of the breather dynamics.
This mechanism, based on the existence of two conservation laws, provides a new perspective over the
 statistical description of condensation processes.
\end{abstract}

\maketitle

Statistical systems exhibiting negative (absolute) temperatures (NTs) often display awkward physical properties, but a
careful analysis shows that their behavior does not contradict any fundamental law of thermodynamics~\cite{Baldovin_2021_review}.
NTs  typically emerge when the total energy is somehow limited from above~\cite{ramsey56}, but they can occur also 
in systems with unbounded energies, provided that the first moment of the density of states in energy remains finite~\cite{machlup75}. 
Lasers are a clear example of the first kind: maximal energy is attained only by setting all the atoms in the excited state.
Vortices in bounded fluids are an example of the second kind~\cite{onsager49,gauthier19,johnstone19}.
Other recent observations of NT states include cold atoms in optical lattices~\cite{braun13} and multimode optical fibers~\cite{baudin23,marques23}.

Energy limitation can also arise indirectly in the presence of additional conservation laws.
The discrete nonlinear Schr\"odinger (DNLS) equation, the subject of this Letter, 
is perhaps the most relevant such model~\cite{kevrekidis09}.
It is used to study many physical phenomena involving propagation of nonlinear waves in discrete media, 
from ultracold gases~\cite{TS,livi06,hennig10} in optical lattices to 
arrays of optical waveguides~\cite{christodoulides88,Lederer2008}. 
In the DNLS model, 
energy and norm (or mass) are conserved. For any mass density, one can construct states of arbitrary
energy, but above a critical energy density (corresponding to a so-called infinite temperature),
the entropy decreases upon increasing energy: the standard signature of a NT. 
As a result, typical states involve the ``condensation" of a finite fraction of energy on a few sites.

Nevertheless, in~\cite{Gradenigo2021_JSTAT,Gradenigo2021_EPJE}, it was proven
that finite-size corrections are so strong that condensation
is inhibited in a strip of width $N^{-1/3}$ ($N$ being the chain length),
above the infinite-temperature line. 
The strip is large enough to justify some numerical results,
such as the observation of ``stationary'' spatially homogeneous states~\cite{Iubini2013_NJP}
and of ``non-Gibbsian'' states~\cite{Flach2018_PRL} (the latter ones detected via a
careful analysis of excursion-time distributions).

The complexity of the underlying regime can be traced back to the emergence of
localized fast rotations known as {\it discrete breathers}, around which a sensible amount
of energy can condense. In the past, breathers have been studied as single dynamical objects,
by investigating the stability of energy clumps~\cite{TS,hennig16,franzosi11REV}.
Here, we aim at characterizing breather dynamics in a statistical-mechanics setup, by singling out the role
of thermodynamic observables.

A statistical description of localization stability has been already produced for breathers interacting with 
a bath of waves 
in terms of a matching condition between breather frequency and the chemical potential of the wave background~\cite{rumpf07,rumpf09,rumpf21}.
Furthermore, tall (i.e. rapidly rotating) breathers have been found to undergo extremely slow 
relaxation phenomena~\cite{ng09,eckmann18,PRL_DNLS}.

So far, this problem was discussed in isolated systems.
In this Letter, we study a setup where mass and energy are exchanged with NT reservoirs,
by introducing an effective grand-canonical measure, which replaces the
non-normalizable standard definition~\cite{Rasmussen2000_PRL}.
As a result, we identify a novel class of extended NT states
and describe the stationary regimes in terms of a properly truncated distribution. 
With the help of an effective Langevin equation
(after including some precautions to get rid of
non-normalizable probabilities~\cite{aghion19,aghion20}),
the steady description is inserted into a dynamical context,
which leads to two predictions: 
(i) condensation progresses (on average) only above a critical mass threshold $c_{max}\sim |T|$,
in agreement with the results in~\cite{rumpf07,rumpf09,rumpf21};
(ii) the lifetime of the associated metastable state is exponentially long (with $|T|$).
The lifetime is lenghtened by the dynamical slowing down of tall breathers~\cite{PRL_DNLS}, 
{\it de facto} eluding the condensation instabilities and making 
NT states physically accessible also in open setups~\cite{kurnosov24,baudin23,Iubini2017_Entropy}.

\paragraph{Grand-canonical description}

In one dimension, the DNLS equation  writes
\begin{equation}
i \dot {z}_n = -2 |z_n|^2z_n - z_{n+1}-z_{n-1} \; ,
\label{eq:dnls}
\end{equation}
where $z_n$ is a complex variable, $n=1,\dots,N$ is the index of the lattice site and suitable boundary conditions are assumed (see below).
The model has two conserved quantities: the total mass $A=\sum_n |z_n|^2$ and the total 
energy~\footnote{Sometime in the literature, the quartic term is preceded by a coefficient $\nu$ quantifying the
nonlinearity~\cite{Rasmussen2000_PRL}. It can be eliminated via a suitable rescaling. 
Its explicit presence implies that the expression of the critical threshold 
$c_{max}$ should be divided by $\nu$ (see below).}
\begin{equation}
 H= \sum_{n=1}^{N} \left( |z_n|^4+z_n^*z_{n+1}+z_nz_{n+1}^* \right) ,
 \label {Hz}
 \end{equation}
The equilibrium behavior of the model can be described in terms of the mass density $a=A/N$ and the energy density $h=H/N$.
Equivalently, the inverse temperature $\beta$ and chemical potential $\mu$~\cite{Rasmussen2000_PRL} can be used
within the ``grand-canonical'' representation. 
Below the infinite-temperature line $h_c=2a^2$, 
homogeneous positive-temperature states exista and are  well described by the Gibbs probability
distribution $P_G(z_n)\sim \exp(-\beta H +\beta\mu A)$. 
Above this line (i.e. for $\beta<0$), the grand-canonical picture is formally ill-defined due to the divergence of $P_G(z_n)$
for arbitrarily large $|z_n|$.

\paragraph{Simplified model}
Close to $\beta=0$,  the hopping energy $H_{int}=\sum_n (z_n z_{n+1}^* + z_n^* z_{n+1})$  can be neglected in Eq.~(\ref{Hz})~\cite{arezzo22}.
The resulting model, called ``C2C''~\cite{JSP_DNLS,Szavits2014_PRL,Gradenigo2021_JSTAT,GIP21,arezzo22,gotti22},
 displays a phase-diagram very similar to that of the DNLS equation.
It can be implemented as a stochastic evolution 
 involving real variables $c_n = |z_n|^2 \ge 0$ in such a way that their sum and the sum of their squares are both 
conserved~\cite{sup_mat}.

In the positive-temperature region, the invariant measure is the product of single-particle distributions
\be
P(c) \propto \exp[-(\beta c^2-m c)]  \quad , 
\label{eq:prob}
\ee
where $m=\beta \mu$.  For $\beta>0$, the probability density is normalizable and thus well defined; spatial density profiles are delocalized.
For $\beta <0$ (and $m<0$), $P(c)$ exhibits a minimum for $c_{max}= m/(2\beta)=\mu/2$, above which it 
diverges. 
Considering the related breather frequency 
$\omega_{max}\simeq 2c_{max}$, the condition writes $\omega_{max}=\mu$
and agrees with~\cite{rumpf07,rumpf09,rumpf21}.
We conjecture and then numerically verify that the regime where $c<c_{max}$ is well described by a
grand-canonical formalism, based on the regularized and factorized 
partition function $Z_r^N = (Z_r)^N$. $Z_r= \int_0^{c_{max}} P(c)\,dc$
can be computed exactly as a function of $m$ and $T=1/\beta$
\be
Z_r = - \sqrt{|T|} \tilde D\left( m\sqrt{|T|}/2 \right)  \,,
\ee  
where 
$\tilde D(x)=e^{-x^2} \int_0^x e^{t^2}\,dt$
is the Dawson function~\cite{abramowitz68,sup_mat}
(from now on, we freely refer to either $\beta$ or $T$, depending on the context).

The average densities $a=\langle A\rangle /N$ and $h=\langle H \rangle /N$ can be 
derived from the grand-canonical relations, 
$a = \partial _m \log(Z_r)$,
$h = \partial_T \log(Z_r)/T^2$ (see~\cite{sup_mat}).
For $|T|\gg 1$, the expressions reduce to
\be
\label{eq:ah}
\begin{cases}
a &=  -\frac{1}{m} +\frac{4\beta}{m^3} -\frac{40\beta^2}{m^5} +o(\beta^2)\\
h &=	  \frac{2}{m^2} -\frac{20\beta}{m^4} + o(\beta)\,
\end{cases}
\ee  
which hold independently of the sign of $\beta$.
Hence, as long as $c< c_{max}$, the stationary properties (including those of spatial profiles) change smoothly while
crossing the infinite-temperature line.

\begin{figure}[ht!]
\includegraphics[width=0.45\textwidth,clip]{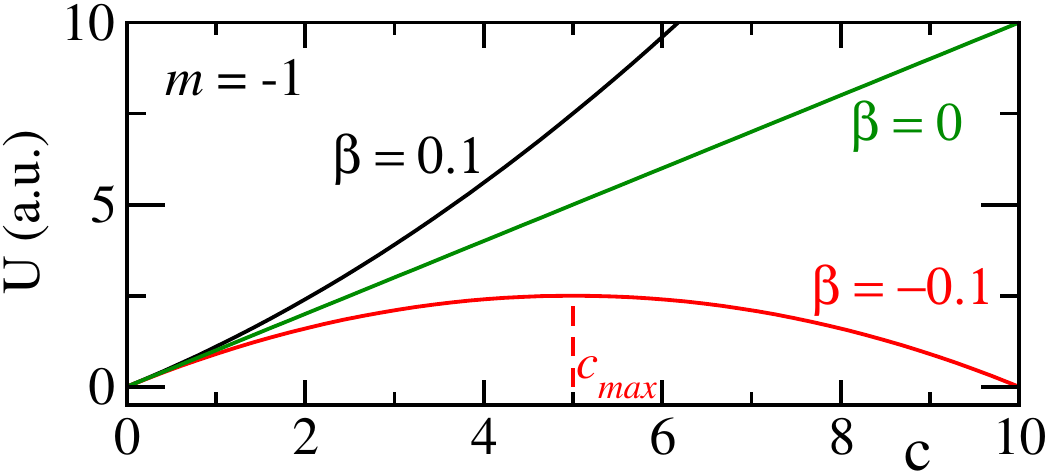}
\caption{Effective potential $U(c)$ when passing from $\beta>0$ to $\beta<0$ for fixed $m=-1$. }
\label{fig:potential}
\end{figure}

One can formally interpret $U(c) \simeq -\ln P(c) = \beta c^2 - mc$ 
as the effective potential (complemented by the condition $U(0) =+\infty$) in
an underlying Langevin process.
A few instances are presented in Fig.~\ref{fig:potential}: $U(c)$ is binding for $\beta > 0$ (see the black curve).
For $\beta =0$, $U(c)$ is still binding provided that $m<0$: see the green straight curve, which corresponds to a
pure exponential distribution.
For $\beta<0$, $U(c)$ is no longer binding, but it is confining for 
$c<c_{max}$ (the red curve in Fig.~\ref{fig:potential} corresponds to $\beta = -0.1$).

In order to validate this formal interpretation, we consider the restriction
of the C2C model to a single triplet $(c_{-},c,c_{+})$, where $c_{-}$ and $c_{+}$ are independent random variables,
both distributed according to Eq.~(\ref{eq:prob}).
It is straightforward to conclude that also the distribution of $c$ is
described by Eq.~(\ref{eq:prob}) for the same $\beta$ and $m$ values.
For NT, $P(c)$ is formally unbounded.
However, consistently with our effective grand-canonical approach, it makes sense,
slightly above the $\beta = 0$ line, to cut the range of allowed $c$ values
at $c_{max}$ (i.e. at a minimum of $P(c)$).

\begin{figure}[ht!]
\includegraphics[width=0.45\textwidth,clip]{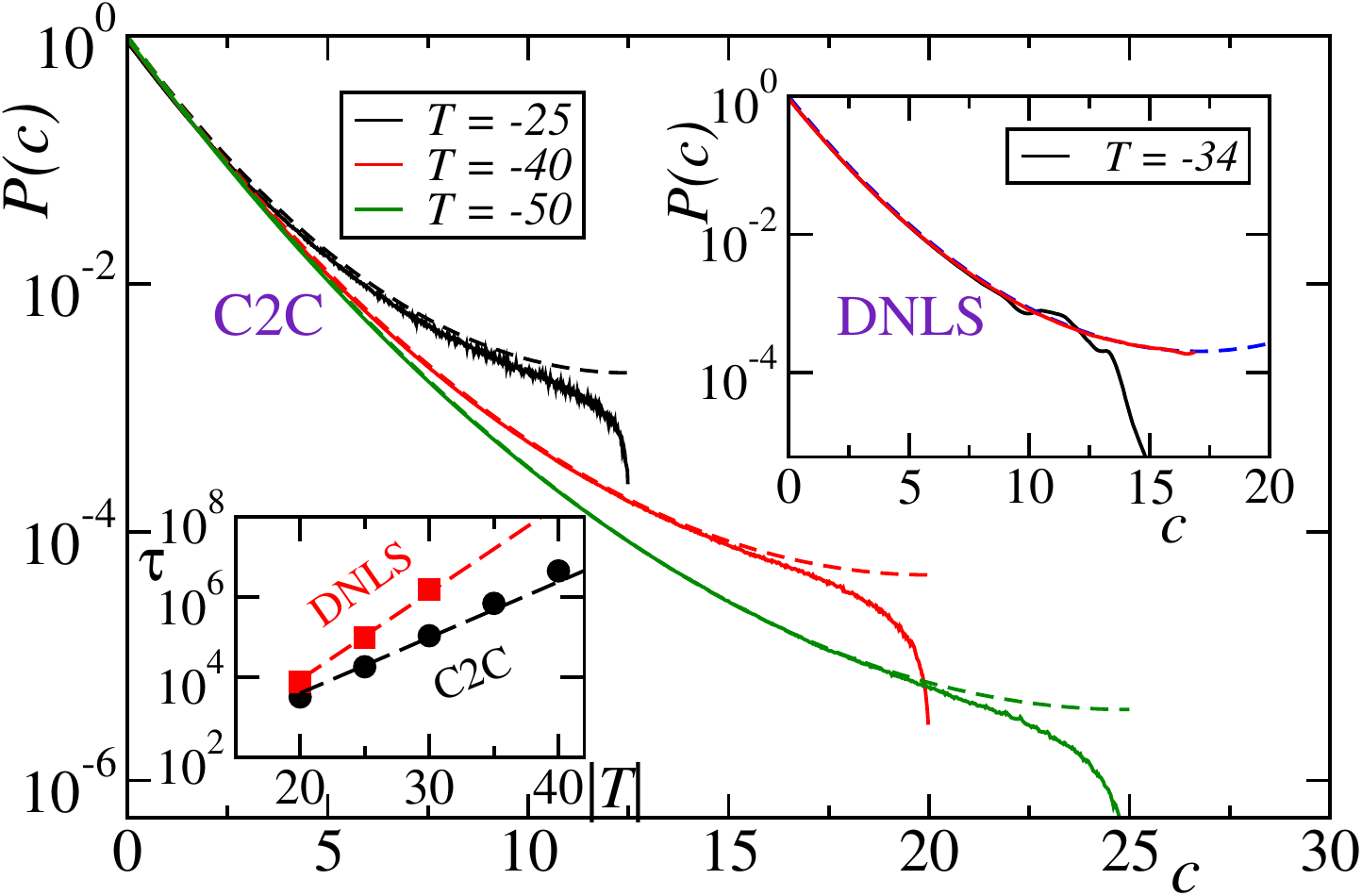}
\caption{Numerical histograms (solid lines) of $c$ values for different negative $T$ for the C2C model. Dashed lines refer to the
corresponding truncated grand-canonical distributions. 
Upper inset:  same analysis for the DNLS chain, Eq.~(\ref{eq:dnlsL}). Black curve refers to bulk sites, the red one to the thermalized site.
Lower inset: first passage time from the saddle {\it vs} $T$  for C2C (full circles) and DNLS (full squares). 
Dashed lines show the theoretical expectations.
}
\label{fig:histofp}
\end{figure}

In Fig.~\ref{fig:histofp} we present the distribution of $c$ values obtained by averaging over 1000 different
trajectories sampled until they reach $c_{max}$ (the initial condition having been 
fixed in $c=1$).
The curves refer to large NT; in all cases the numerical histograms closely
follow the theoretical expectation (i.e. the truncated form of Eq.~(\ref{eq:prob})), 
confirming the validity of the effective grand-canonical approach. 

\paragraph{Effective Langevin description}
Here, we approximate the model dynamics with the stochastic equation
\be
   \dot c = F(c) + \sqrt{2 D(c)}\xi \; ,
\label{eq:lang}
\ee
where the deterministic drift $F(c)$ and the diffusion coefficient $D(c)$ ($\xi$ is a white noise with unit variance)
are empirically defined from numerical simulations
\be
F(c) = \langle c'\rangle - c  \quad , \quad 2D(c) = \langle (c'-c)^2 \rangle  \; .
\ee
Here $c'$ is the iterate of $c$ as obtained by implementing the C2C model.
The angular brackets denote an average over all possible realizations of
the thermal baths and implementations of the C2C rule.

\begin{figure}[ht!]
\includegraphics[width=0.45\textwidth,clip]{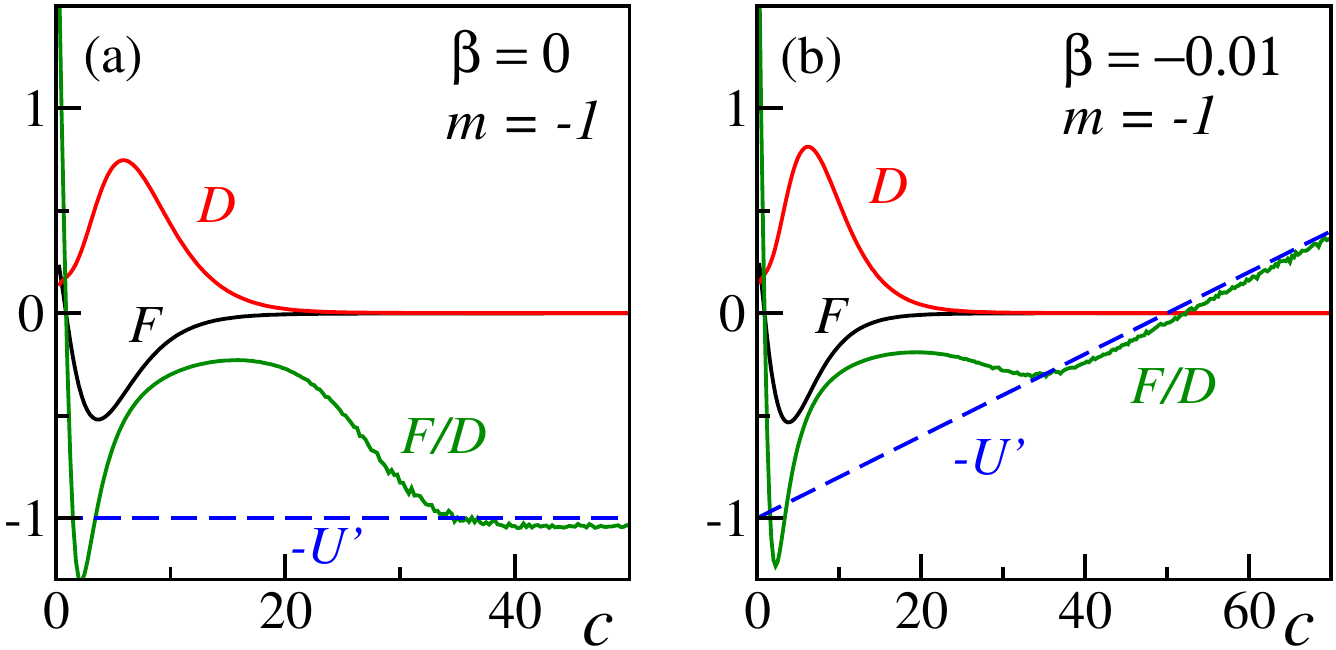}
\caption{Velocity field $F(c)$ and diffusion $D(c)$ obtained from numerical simulations of the C2C model for $\beta=0$ (a) and $\beta=-0.01$ (b). In both cases, the corresponding ratio $F(c)/D(c)$ is compared with the expected force $-U'(c)$. }
\label{fig:fd}
\end{figure}

In Fig.~\ref{fig:fd}, we see that both for $\beta=0$ and $-0.01$,
$F$ is negative for (not too) small $c$ values, indicating that small breathers are stable.
$F/D$~\cite{gardiner85} provides a better representation.  
For $\beta=0$, we see that the drift is still negative at large  $c$ values (see panel (a))
in agreement with the expectation that breathers are eventually absorbed.
For $\beta=-0.01$, the drift crosses zero at $c=50$, becoming positive (see the green line in panel (b)) 
in agreement with the theoretical expectation 
$c_{max} = mT/2$ ($m=-1$).
On a yet more quantitative level, the asymptotic linear growth of $F/D$ matches the linear
dependence of $-U'(c) = 2\beta c -m$, thus showing that the Langevin equation reproduces
the behavior of the C2C model for large $c$ values.
In particular, only breathers with
 $c>c_{max}$ tend to explode, 
or, equivalently, represent seeds of condensation.

Once $F$ and $D$ have been determined, one can estimate the lifetime of the metastable state via
Kramers formula~\cite{risken_book} for the escape through a barrier located in $c_{max}$,
$\tau \approx \exp(\Delta U)/\sqrt{D_0D_1}$, where $\Delta U$ is the barrier height,
while $D_0\equiv D(0)$ and $D_1\equiv D(c_{max})$ are the diffusion coefficients 
in the bottom of the valley and at the saddle, respectively.
Since $\Delta U = m^2|T|/4$, $\tau \approx \exp(\lambda |T|)$, with $\lambda = m^2/4$.
Numerical simulations performed for $m=-1$ and $T\in [-50,-25]$ yield $\lambda \approx 0.365$ to be compared
with the theoretical prediction 0.25.
Large part of the discrepancy can be attributed to the indirect dependence of $D_1$ on $T$.
In fact, as visible in both panels of Fig.~\ref{fig:fd}, $D$ decreases with $c$, 
 while $c_{max}=-T/2$.
 The effective dependence follows a power law~\cite{JSP_DNLS,JSTAT_mmc} and thus it does not 
affect the asymptotic scaling
behavior; however, in the range of temperatures here explored,
  $D_1$ contributes an increase of
   $\lambda$ by about 0.07, bringing the theoretical prediction
closer to the observed value (0.32 vs 0.365).
The remaining discrepancy is presumably 
due to the imperfect description of the stochastic dynamics in terms of a continuous nonlinear Langevin process.

\paragraph{DNLS dynamics}

Given the above mentioned continuity with the high-temperature region, 
it makes sense to extend the scheme proposed in~\cite{iubini13off} to model the
interaction of a DNLS chain with NT reservoirs 
\be
\dot z_n\! = \!(\gamma_n-i)(-2|z_n|^2 z_n\! -\!z_{n+1}\! -\!z_{n-1}) +\gamma_n\mu z_n + \sqrt{\gamma_n T}\xi\,
\label{eq:dnlsL}
\ee
where $\gamma_n=\delta_{n1}\gamma $ is the coupling strength and $\xi$ a complex Gaussian noise with zero mean and unit variance.
Periodic boundary conditions ($z_{N+1}=z_1$) are assumed.
While $\gamma$ is positive for $T >0$, it changes sign at NT.
This property is in accord  with the general symmetries expected for NT
reservoirs~\cite{Baldovin_2021_review};
$\gamma<0$ indicates that the sign of nonconservative forces is reversed with respect
to the $\beta>0$ case. 
If we neglect the nearest-neighbor interactions of the thermostatted site, the evolution
equation of the mass $c= |z_1|^2$ writes
$\dot c = -2\gamma(2 c^2 - \mu)c + z^*\xi + z \xi^*$, a form similar to the Langevin Eq.~(\ref{eq:lang}), with
the difference that this equation follows from a precise thermal-bath definition~\cite{iubini13off}
instead of being a mere conjecture.
It is also interesting to notice that the effective force
$\tilde F(c) = 2\gamma c \left(\mu-2c^2  \right)$ predicts a destabilization
for $c^2\geq c_{max}$, consistent with the single-particle probability $P(c)$, Eq.~(\ref{eq:prob}). 

\begin{figure}[ht!]
\includegraphics[width=0.5\textwidth]{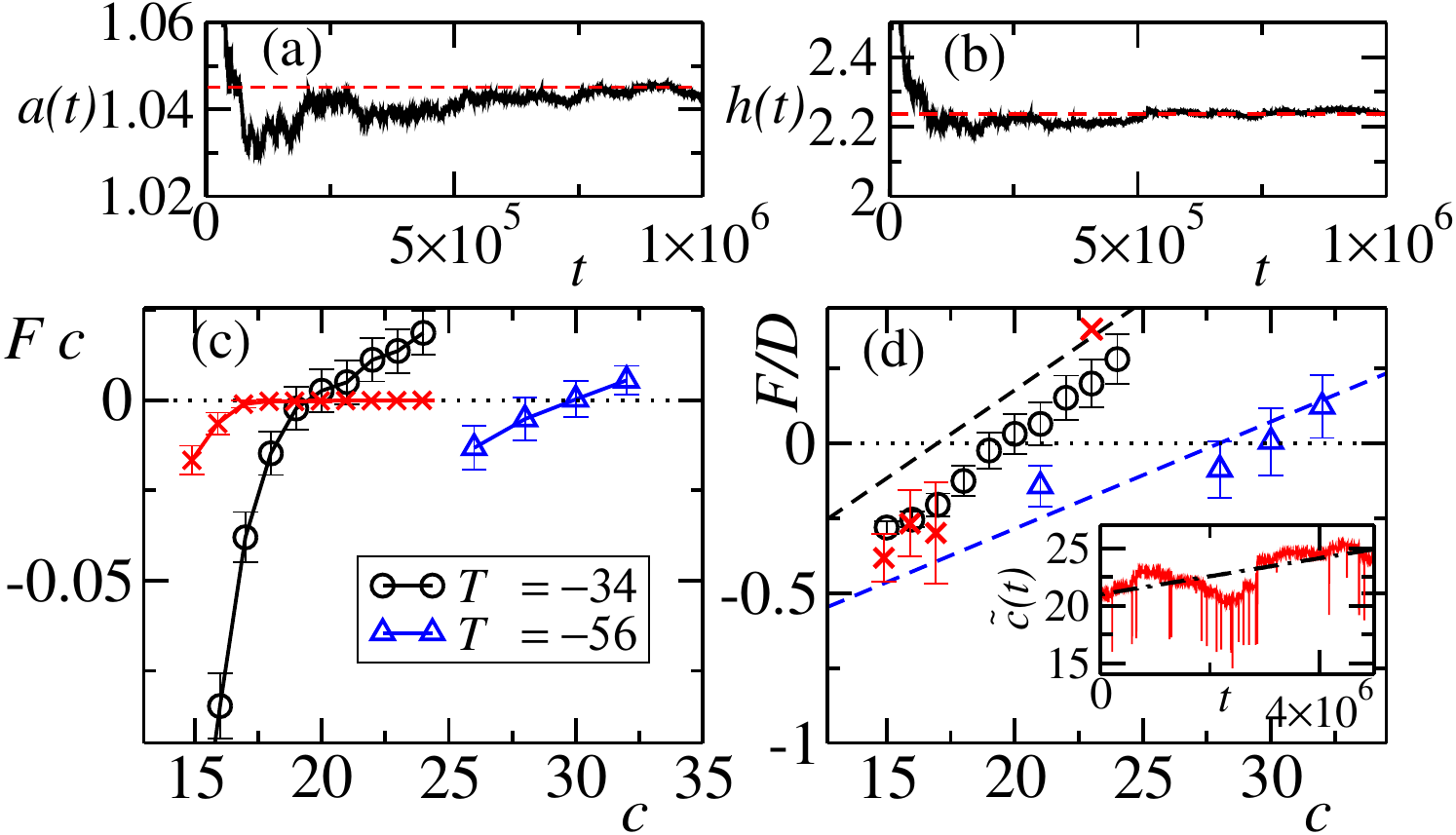}
\caption{
 (a,b) Relaxation to pseudo-equilibrium of a DNLS equation interacting with a generalized Langevin reservoir with $\beta=-0.01$, $m=-1$ and $\gamma=-0.02$: evolution of average mass $a(t)$ and average energy $h(t)$. Red dashed lines refer to the analytical prediction in~(\ref{eq:ah}). 
(c) Effective force $F$ (scaled by $c$ for clarity) acting on a breather with initial mass $c$
for $m=-1$ and different $T$. 
Open symbols refer to the SDNLS model; red crosses to
the full DNLS for $T=-34$. 
 (d) Ratio $F/D$ compared to theoretical values of $-U'(c)$ (dashed lines). Same symbols as in (c).
 Inset: DNLS evolution of $\tilde c(t)$ for $c=21$. The dot-dashed line shows the average trend excluding downward bursts.  
  }
\label{fig:stocdnls_force}
\end{figure}

Fig.~\ref{fig:stocdnls_force} (a,b) shows the averages of $a$ and $h$ in a DNLS chain of length $N=20$,
thermostatted for $T=-100$, $m=-1$ and $\gamma=-0.02$
(in this regime $\tau > 10^9$ is much longer than the integration time).
They are in good quantitative agreement with the truncated grand-canonical prediction Eq.~(\ref{eq:ah})
(dashed lines), as also confirmed by the direct extraction of mass distributions, see the upper inset of Fig.~\ref{fig:histofp}. 
This correspondence is not {\it a priori} obvious, since the theory neglects $H_{int}$.
In the metastable $\beta<0$ region of interest for this work, $H_{int}$ is on average negligible~\cite{sup_mat}.

\paragraph{DNLS (frozen) instability} 

We now study the evolution of discrete breathers for different amplitudes $|z^2|$. 
At variance with the C2C model, here thermostats induce strong boundary effects.
As previously shown in positive-temperature simulations~\cite{PPI2022_JSTAT}, the high frequencies
naturally generated by the heat bath spuriously affect the DNLS dynamics (in particular the
breather stability). Hence, Eq.~(\ref{eq:dnlsL}) cannot account for the true dynamics of the background.
Nevertheless, Eq.~(\ref{eq:lang}) provides again a good effective description.
By determining the drift $F$ and diffusion $D$ of breathers of mass $|z|^2=c$ 
(located in the antipodal position with respect to the thermal bath) for different
chain lengths, we have verified that a length $N=6$ suffices to kill unphysical high
frequencies.

To improve numerical accuracy~\cite{PPI2022_JSTAT}, we monitor the  square root of breather energy $\tilde c(t)=\{\,|z_{n_0}|^4+ [z_{n_0}^*(z_{n_0+ 1}+z_{n_0-1})+ c.c.]\,\}^{1/2}$, (dimensionally equivalent to the local mass), and compute $F(c)=\langle d \tilde c /dt \rangle$. Analogously, $D(c)=Var[\,\tilde c(t)\,]/2t$, where 
$Var(\cdot)$ denotes the variance of the signal.
The red crosses reported in Fig.~\ref{fig:stocdnls_force} show that the drift is negative for small
$c$-values, indicating that such breathers are stable, consistently with the numerical evidence of pseudo stationary regimes.
However, at variance with the stochastic C2C model, here $F(c)$ becomes indistinguishable from zero above $c\approx 17$.
This slowing down is a manifestation of the adiabatic invariant unveiled 
in~\cite{PRL_DNLS,PPI2022_JSTAT} in the positive-temperature region.

In order to bypass this obstacle, we have also analyzed a stochastic version of the DNLS equation (SDNLS), where the 
adiabatic invariant is absent.
It amounts to including the sporadic update of
the phase of a randomly chosen $z_n$ so as to leave the local interaction energy unchanged~\cite{sup_mat,iubini19}.
The SDNLS exhibits the same stationary properties as the DNLS model, but is characterized 
by a substantially faster breather dynamics.

Simulations of the SDNLS chain show that the drift is now significantly different 
from 0 and quantitatively measurable (see circles and triangles in Fig.~\ref{fig:stocdnls_force}(c)).
In particular, one can identify a crossing point above which the drift is positive.
Also the diffusion coefficient can be estimated; the resulting numerical values of $F/D$ are
presented in Fig.~\ref{fig:stocdnls_force}(d).
They are relatively close to the theoretical predictions (see the dashed lines);
the main difference is a rightward shift which implies that
the numerically estimated saddle point $\tilde c_{max}$ is slightly larger than the expected value
(on the basis of the C2C model). Anyway, 
the deviation $\tilde c_{max} - c_{max}$ does not increase upon increasing $|T|$~\cite{sup_mat}.

A fully quantitative analysis of the deterministic DNLS equation is not doable. However, we have made our best efforts to
determine $F/D$ in a single point in the presence of an adiabatic invariant.
We have performed a single long numerical simulation for $T=-34$ and an
initial breather mass $c=21 >c_{max}=17$ for $N=12$: the evolution  of $\tilde c(t)$
is reported in the inset of Fig.~\ref{fig:stocdnls_force}(d).
It reveals large 
 bursts and less visible but relevant blobs which all hinder a
quantitative analysis, since these localized instabilities do not typically contribute to the overall growth and 
diffusion~\footnote{They signal the emergence of ``dimer'' bound states when the energy temporarily spreads over two adjacent sites}
and yet strongly affect numerical estimates.
After a lengthy manual removal of the most relevant bursts, we have obtained a cleaner curve
~\cite{sup_mat}, whose analysis leads to $F/D \approx 0.38$,
in semi-quantitative agreement with the theoretical expectation (see the uppermost red cross in Fig.~\ref{fig:stocdnls_force}(d)).

Finally, we consider the escape time out of the DNLS  metastable regime.
The major difference with the C2C model is the dependence of the diffusion coefficient $D$ on the mass
contained in the breather. As shown in~\cite{PRL_DNLS,PPI2022_JSTAT}, the adiabatic invariant induces an exponential
decrease, $D \approx \exp(-\alpha c)$, with $\alpha\approx 1$. Hence, $D_1 \simeq \exp(-\alpha mT/2)$, which, inserted into
the Kramers formula, enhances the exponential growth rate  of the lifetime, to 
$\lambda \approx m^2/4 + \alpha |m|/4$. This implies (for $m=-1$), $\lambda \approx 0.5$.
Direct numerical simulations performed for
$N=7$  yield the results reported in Fig.~\ref{fig:histofp};
an exponential fit gives a rate 0.53 in good agreement with the theoretical expectation.

\paragraph{Conclusions}
We have revisited the stability of extended states by implementing a grand-canonical approach and
introducing suitable thermal reservoirs, which operate at large NTs.
Our main achievement is the demonstration of metastable states characterized by
exponentially long lifetimes. This result follows from the existence of a 
temperature-dependent critical amplitude, which separates
stable from unstable condensation peaks (DNLS breathers).
In the DNLS model, the exponential growth is amplified by the fast decrease of the
diffusion coefficient with the breather height.

In this Letter, we focused on a setup where the local onsite energy $\epsilon$ is a quadratic function of the mass
$c$, but the formalism can be applied to contexts where $\epsilon(c) > 0$ (see, e.g.,
purely stochastic condensation models~\cite{Szavits2014_PRL,Szavits2014_JPA}): in such cases
$c_{max}$ solves $\beta \epsilon'(c_{max})=m$.
Finally, we have assumed the presence of a single breather; in general, multiple breathers
are expected to arise in the NT region, each accompanied by an additional quasi-conserved quantity.
It is conceivable that a generalized Gibbs ensemble should be implemented, analogous to what done 
in the context of integrable systems~\cite{vidmar16,spohn20}.
This generalization is, however, not straightforward; it requires further work.

\begin{acknowledgments}
We thank P. Politi for fruitful discussions.
We acknowledge support from
the project ``Breakdown of ergodicity in classical and quantum many-body systems'' (BECQuMB) funded by the Italian Ministry for University and Research (MUR) PRIN2022 grant No. 20222BHC9Z.
\end{acknowledgments}

%

\end{document}


\renewcommand{\thepage}{S\arabic{page}}  
\renewcommand{\thesection}{S\arabic{section}}   
\renewcommand{\thetable}{S\arabic{table}}   
\renewcommand{\thefigure}{S\arabic{figure}}
\renewcommand{\theequation}{S\arabic{equation}}
\renewcommand{\bibnumfmt}[1]{[S#1]}
\renewcommand{\citenumfont}[1]{S#1}

\title{Supplemental material for\\
\vspace{0.5 cm	}
Effective grand-canonical description of condensation in negative-temperature regimes}

\author{Stefano Iubini}

\author{Antonio Politi}

\maketitle

\subsection{C2C stochastic model}
\begin{figure}[ht!]
\includegraphics[width=0.4\textwidth]{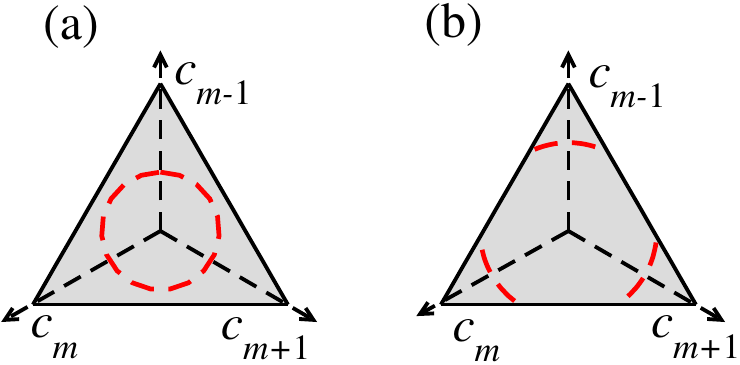}
\caption{Available configurations of a C2C triplet (red dashed line) lie either on a full circle (panel (a)) or on three distinct arcs (panel (b)).}
\label{fig:c2c}
\end{figure}

The model is characterized by the presence of two conserved quantities: $H =\sum_n c_n^2$ and $A =\sum_n c_n$, with $c_n\ge0$.
Dynamical simulations are performed as a sequence of Monte Carlo moves~\cite{JSP_DNLS,Szavits2014_JPA,JSTAT_mmc,GIP21}.
A triplet of consecutive sites is randomly selected, and the corresponding variables $(c_{m-1},c_m,c_{m+1})$  are
updated to $(c'_{m-1},c'_m,c'_{m+1})$  
under the constraint that
their sum $A_m = c'_{m-1}+c'_m+c'_{m+1}$ and the sum of their squares 
$H_m= (c')^2_{m-1}+(c')^2_m+(c')^2_{m+1}$ are both conserved.
Hence, the admissible configurations lie along the circle resulting from the intersection between the surface 
of a sphere of radius $\sqrt{H_m}$ and the plane determined by the value $A_m$. 
The new configuration is randomly selected by attributing an equal angular weight to all
points sitting in the octant characterized by three positive variables.
If $H_m \ge A_m^2/2$, the circle is fully contained in the octant, see Fig.~\ref{fig:c2c}(a); otherwise,
the positivity constraint is satisfied only in three distinct arcs, see Fig.~\ref{fig:c2c}(b): in this case, 
the updating procedure is restricted to the arc containing the initial configuration. 
It can be seen that this protocol satisfies detailed balance.

\subsection{Regularized grand-canonical description of metastable states}

We provide an analytical description of negative-temperature (NT) metastable states based on the single-particle distribution  $P(c)=P(0)\exp[-(\beta c^2-mc)]$.

The regularized partition function writes explicitly (we set $P(0)=1$ with no loss of generality)
\be
Z_r =\int_0^{c_{max}} P(c)\,dc= \int_0^{c_{max}}  e^{-\beta c^2 +mc} \,dc\,,
\ee
where $c_{max}=m/(2\beta)$ is the cutoff value. 

The function $Z_r$ can be rewritten as 
\be
Z_r = \frac{e^{\frac{m^2}{4\beta}}}{\sqrt{\beta}} \int_{-\frac{m}{2\sqrt{\beta}}}^0 e^{-y^2}\,dy\,,
\ee
where the integral is meant along the imaginary axis ($\beta<0$).
We incidentally remark the formal similarity with the partition function $Z_0$ in the positive-temperature region~\cite{gotti22}
\be 
Z_0 = \frac{e^{\frac{m^2}{4\beta}}}{\sqrt{\beta}} \int_{-\frac{m}{2\sqrt{\beta}}}^\infty e^{-y^2}\,dy\,.
\ee
In terms of the error function, one has
\be 
Z_r = \frac{e^{\frac{m^2}{4\beta}}}{\sqrt{\beta}}\frac{\sqrt{\pi}}{2}\erf\left(\frac{m}{2\sqrt{\beta}}\right)\,.
\ee
In order to proceed, we specify $\beta=-|\beta|$. Accordingly,
\be
Z_r = -\frac{1}{2}\sqrt{\frac{\pi}{|\beta|}} e^{-\frac{m^2}{4|\beta|}}\erfi\left( \frac{m}{2\sqrt{|\beta|}}\right)\,,
\ee
where $\erfi(x)= i\,\erf(-ix)$ denotes the so-called imaginary error function~\footnote{Notice that the function $\erfi(x)$ is real-valued.}. Upon expressing
$\erfi(x)=2/\sqrt{\pi}\, e^{x^2} D(x)$, where $\tilde D(x)$ is the Dawson function~\cite{abramowitz68}
\be 
\label{eq:daw}
\tilde D(x)=e^{-x^2} \int_0^x e^{t^2}\,dt\,,
\ee
we finally obtain
\be
Z_r = -\frac{1}{\sqrt{|\beta|}} \tilde D\left( \frac{m}{2\sqrt{|\beta|}} \right).
\ee

For given parameters $(\beta\leq 0,m\leq 0)$ in the metastable regime, the corresponding average densities follow from
\be
\begin{cases}
a &= Z_c^{-1}\int_0^{c_{max}} c\, e^{-\beta c^2 +mc} \,dc\, = \partial _m \log(Z_r)\\
h &= Z_c^{-1}\int_0^{c_{max}} c^2\, e^{-\beta c^2 +mc} \,dc\, =-\partial_\beta \log(Z_r	)\,.
\end{cases}
\ee
After some algebra, using the property  $d\tilde D(x)/dx = 1-2x \tilde D(x)$, we obtain
\be
\label{eq:ah}
\begin{cases}
a &= \frac{1}{2\sqrt{|\beta|} \tilde D\left( \frac{m}{2\sqrt{|\beta|}} \right)} - \frac{m}{2|\beta|} \\
h &= -\frac{1}{2|\beta|} (ma +1)\,.
\end{cases}
\ee


Close to $\beta=0^-$ and for finite $m$, relations in  Eq.~(\ref{eq:ah}) can be approximated  by expanding 
$\tilde D(x)$ for large $x$,
\be
\tilde D(x)\simeq \frac{1}{2x} + \frac{1}{4x^3} +\frac{3}{8x^5} +\frac{15}{16x^7} +O(x^{-9})\,,\quad |x|\gg 1 .
\ee
We obtain
\be
\begin{cases}
a\simeq -\frac{1}{m}-\frac{4|\beta|}{m^3}-\frac{40\beta^2}{m^5}\\
h \simeq \frac{2}{m^2} +\frac{20|\beta|}{m^4} \qquad (|\beta|\ll 1)\,.
\end{cases}
\ee

By including the expansions previously obtained in the case $\beta>0$~\cite{gotti22}, mass and energy densities read
\be
a\simeq\begin{cases}
-\frac{1}{m} +\frac{4\beta}{m^3} -\frac{40\beta^2}{m^5} \quad &\beta\geq 0 \\
-\frac{1}{m}-\frac{4|\beta|}{m^3}-\frac{40\beta^2}{m^5} \quad & \beta\leq 0
\end{cases}
\ee

\be
h\simeq\begin{cases}
\frac{2}{m^2} -\frac{20\beta}{m^4} \quad &\beta\geq 0 \\
\frac{2}{m^2} + \frac{20|\beta|}{m^4} \quad & \beta\leq 0\,.
\end{cases}
\ee

Therefore, regardless the sign of $\beta$,

\be
\label{eq:ah+-}
\begin{cases}
a &\simeq  -\frac{1}{m} +\frac{4\beta}{m^3} -\frac{40\beta^2}{m^5} \\
h &\simeq  \frac{2}{m^2} -\frac{20\beta}{m^4}\qquad (|\beta|\ll 1).
\end{cases}
\ee

Eq.~(\ref{eq:ah+-}) clarifies, perturbatively, that there are no discontinuities in $(a,h)$ around $\beta=0$. As a result, the transition from 
positive-temperature states to negative-temperature metastable states is smooth.  

\subsection{Computational analysis of the DNLS model}

Langevin equation for DNLS models, Eq.~(8), has been integrated numerically according to a standard fourth-order Runge-Kutta algorithm~\cite{press2007numerical}. 
A sufficiently small time step $\delta t$ even down to $10^{-5}$ units has been chosen in order to follow fast rotations of breather evolution.
Data in Fig. 4 (c-d) (except for the uppermost red cross, see below) are obtained from averages over $10^5$ independent samples prepared by superposing a peak with initial mass $|z|^2=c$ on 
a NT pre-thermalized background. Background thermalization is realized by evolving Eq.~(8) for times of order $10^3$ units starting from 
an infinite-temperature initial condition (exponential distribution of $|z_n|$ and uniform random phases $\arg{(z_n)}$~\cite{Rasmussen2000_PRL}).

\subsubsection{DNLS simulations}

Stability studies of large-amplitude states in the DNLS equation reported in Fig. 4 (c-d) require to analyze the
extremely slow evolution of breather states interacting with a NT environment.
%
%
On the other hand,  the direct application of the stochastic Langevin force defined as in Eq.~(8) 
to a breather site
destroys the adiabatic
invariant accompanying large breathers~\cite{PPI2022_JSTAT}.  As a result, the thermostatted dynamics
is much faster than that of the
deterministic DNLS model.
For this reason, it is necessary to consider not too short systems where the breather sits at a certain distance form the thermostatted site,
%
 as discussed in
Ref.~\cite{PRL_DNLS}, where breather relaxation was investigated at positive temperatures.
In fact, here we have numerically integrated chains with at least 6 sites (with periodic boundary conditions) and analyzed
the behavior of those sites at a distance larger than 2 sites from a thermostat. 

At NT, an additional problem is present: the relatively frequent emergence of breathers 
at the thermostatted site (again a consequence of the missing adiabatic invariant).
Instead of removing, a posteriori, these ``artificial" events, 
we have decided to inhibit them, by reflecting back the amplitude $z_n$
whenever the local mass becomes larger than $c_{max}$. In practice, at each time step, if $|z_n|^2>c_{max}$ (on the
thermostatted site), then $z_n	 \to z_n c_{max}/|z_n|^2$.
We have verified that this artifice does neither affect the mass probability density, nor
prevents the birth of unstable breathers in the bulk.

\subsubsection{Stochastic DNLS (SDNLS) equation}

We have applied the same methodology also for the 
 evolution of the SDNLS model, which is obtained from Eq.(8) by adding suitable updates of the phase $\arg{(z_k)}$ on  randomly chosen  sites $k$ according to Ref.~\cite{iubini19}.
Simple algebra shows that, generically, there exists one and only one solution for the phase update  which differs from the current phase value  and allows the conservation of local energy $|z_{k}|^4+ [z_{k}^*(z_{k+ 1}+z_{k-1})+ c.c.]$, in addition to local mass $|z_k|^2$.
Phase updates are performed at random times on all lattice sites not directly thermalized. Time separations $\hat{t}$ between two consecutive updates on each site are 
independent and identically distributed variables extracted from a Poissonian distribution $P(\hat t)\simeq \exp(-\hat\gamma \hat t)$. We have verified that 
$\hat \gamma=10$ provides a sufficiently fast convergence to stationary states, both for delocalized states and for localized ones. The presence of the external ``phase noise'' here considered 
breaks down the Hamiltonian structure of the original deterministic DNLS model, thereby determining the destruction of adiabatic invariants (AI)s in the SDNLS equation, see also~\cite{PPI2022_JSTAT}  for details.

\begin{figure}[ht!]
\includegraphics[width=0.5\textwidth]{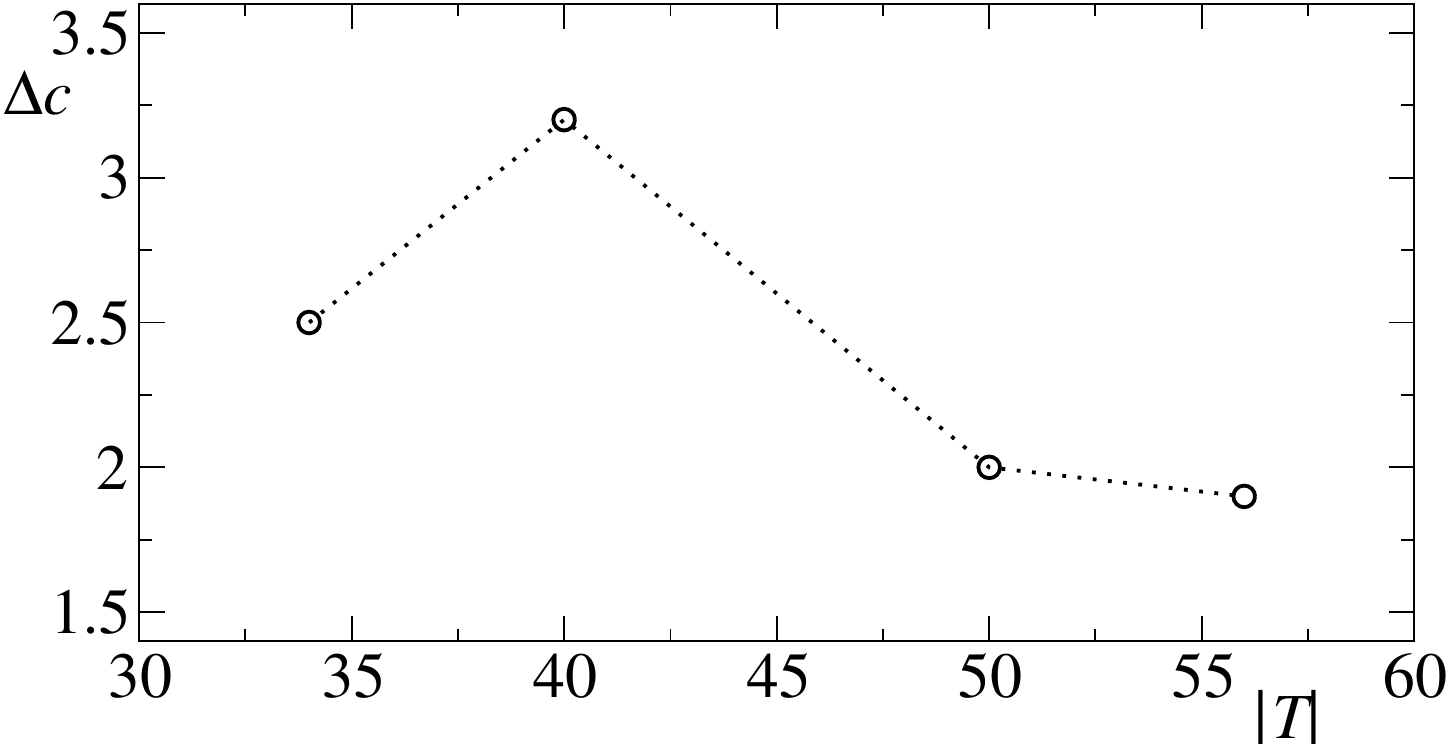}
\caption{Deviation of the numerically estimated saddle point $\tilde c_{max}$ with respect to $c_{max}$ for the SDNLS model. Data refer to the setup of Fig. 4(c).}
\label{fig:delta}
\end{figure}

Because of the absence of AIs, simulations of the SDNLS model can be ``easily" extended to relatively large breathers. The position of the saddle of the potential obtained from Fig.~4(c) does not correspond exactly with the theoretical prediction. Nevertheless, the
deviation $\Delta c=\tilde c_{max}-c_{max}$ does not grow with $|T|$, as clearly visible in Fig.~\ref{fig:delta}.
The scaling behavior of the lifetime of the extended state is thus confirmed.

\subsubsection{Breather dynamics}

For large breather amplitudes, the dynamics of the deterministic DNLS  is so slow and time-correlations are so relevant, that it is
not possible to collect enough statistics for an accurate automatic estimate of the drift and diffusion coefficient.
For this reason, we have processed manually a single very long trajectory, by preliminarly removing several ``local" bursts.
As shown at positive temperatures for unidirectonal coupling~\cite{PPI2022_JSTAT}, the expression of the AI is similar to
but different from the local energy: some fluctuations are not manifestations of true variations but just
the consequence of an approximate knowledge of the AI. In fact, a close look at $\tilde c(t)$ shows localized
bursts which terminate by returning to the initial value; such bursts my last a few thousands of time units.
However, sometimes the bursts terminate by exhibiting jumps either up or down: 
they are a manifestation of persistent fluctuations.
In the absence of a quantitative theory, we have manually removed the largest bumps.

\begin{figure}[ht!]
\includegraphics[width=0.5\textwidth]{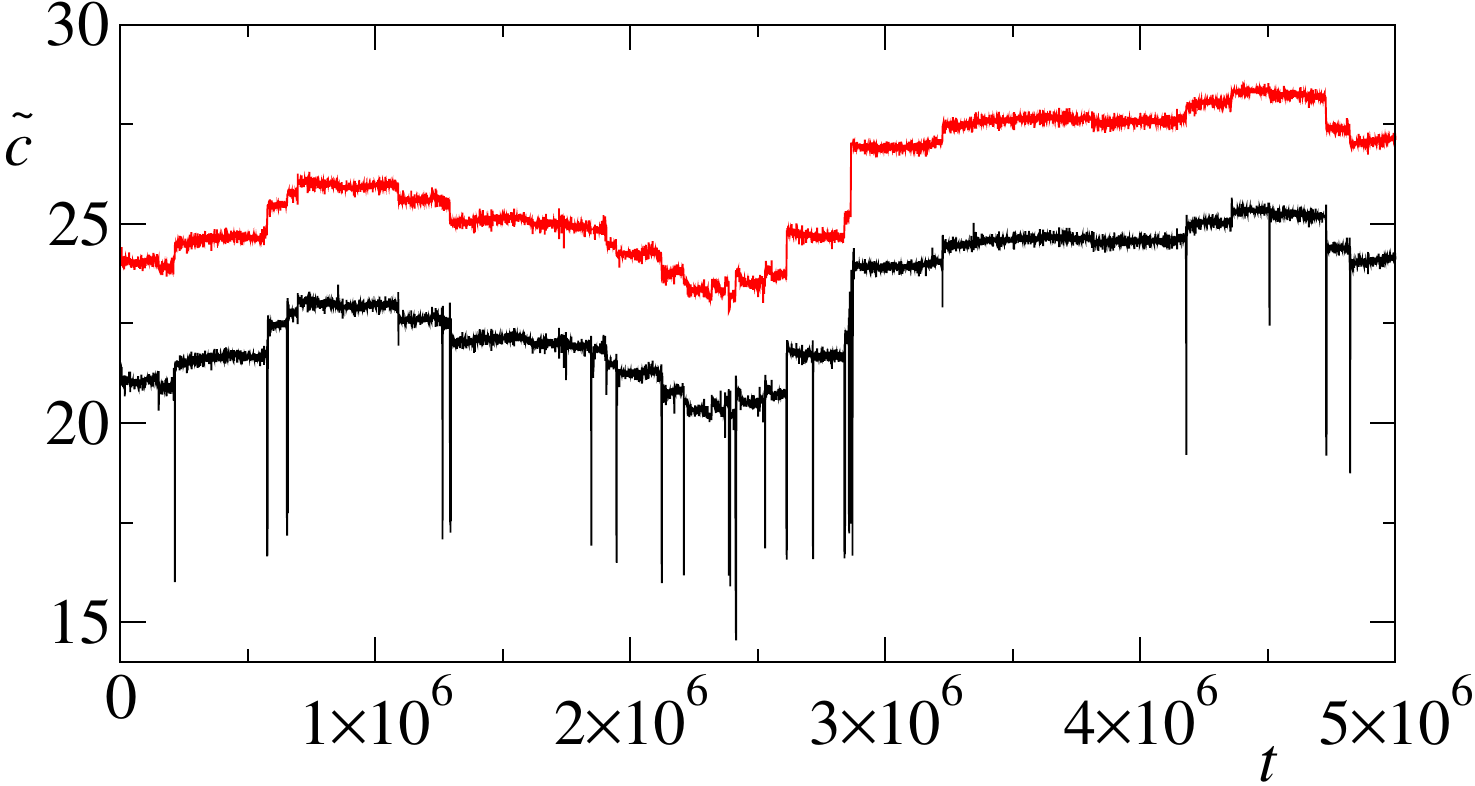}
\caption{Black line: evolution of the mass-like quantity $\tilde c$ for the setup of Fig. 4(d), $m=-1$, $T=-34$, $c(0)=21$. 
Red line: the same trajectory (vertically shifted for clarity) after removal of the bursts.}
\label{fig:traj}
\end{figure}

The resulting filtered trajectory is shown in Fig.~\ref{fig:traj} (see the red curve), where it can be compared with the
original time series). This trajectory has been then split 
into 20 equal-length blocks, to determine the drift and diffusion coefficients by averaging over all the blocks. 
The resulting $F/D$ ratio is reported in the inset of Fig. 4(d) (see the uppermost red cross).

\subsection{DNLS hopping energy in the NT regime} 

Moving above the $\beta=0$ line, the hopping term in the DNLS Hamiltonian  is expected to take a positive value.
Here we quantify its contribution with respect to the value of the local nonlinear energy $|z_n|^4$ for $T=1/\beta=-34$ and $m=-1$.
Two different regimes will be separately analyzed: metastable uniform states and  localized states.

{\it Metastable regime (delocalized) --}
NT Langevin dynamics is implemented according to Eq.~(8), starting from an  infinite-temperature initial condition with $P(|z_n(0)|^2)\sim e^{-|z_n(0)|^2}$.
Metastable trajectories are sampled for $40$ time units after a transient of $2\times 10^3$ units. Reflection algorithm (see above) is implemented on 
the thermalized site in order to avoid the growth of unstable peaks in the whole chain  during the simulation period.

We separately measure local and interaction energies, namely,
\begin{eqnarray}
h^{(loc)} &=& 1/N  \sum_n \langle |z_n|^4 \rangle \\
h^{(int)} &=& 1/N \sum_n \langle z^*_n(z_{n+1}+z_{n-1}) +c.c. \rangle\,,
\end{eqnarray}
with $h^{(loc)}+h^{(int)}=h$. Table~\ref{tab:cfr} summarizes the results. In detail, we obtain a ratio $h_{int}/h_{loc}\simeq 2.5\%$.



\begin{table}[ht!]
\begin{tabular}{c|c|c|c|c}
     &   $a$ & $h^{(loc)}$  & $h^{(int)}$  &  $h$ \\
     \hline
DNLS   &   $1.17$ & $2.98$  & $0.076$  &  $ 3.06 $ \\
\hline
theory  &   $1.18$ & $3.20$  & $0$  &  $3.20$
\end{tabular}
\caption{Mass and energy densities for $T=-34$ and $m=-1$.  DNLS simulations (first row) are performed on a chain with $N=6$ sites and periodic
boundary conditions. A sample of $4\times 10^4$ independent trajectories is considered. $\gamma=-4\times 10^{-2}$. The second row shows the
corresponding densities obtained analytically form Eq.~(\ref{eq:ah}).}
\label{tab:cfr}
\end{table}

{\it Localized states--} In a similar setup, one can quantify the contribution of interaction energy of a breather with mass $c$ that interacts with a stationary
background at negative temperature. For a breather located on site $k$ we define $h^{(int)}_k = \langle z^*_k(z_{k+1}+z_{k-1}) +c.c. \rangle $
 and $ h^{(loc)}_k =  \langle |z_k|^4 \rangle$
For the same parameters ($T=-34$ and $m=-1$), we report in Tab.~\ref{tab:br_int} the results obtained for three different breather masses.

\begin{table}[ht!]
\begin{tabular}{c|c|c}
$c$        & $h^{(loc)}_k$ & $h^{(int)}_k$  \\
\hline
15      &   $220$    & $2.46$  \\
\hline
19		&   $360$  &  $2.32$  \\
\hline
23		&   $530$  & $2.26$  
\end{tabular}
\caption{Local and interaction energies for a breather with mass $b$. Same parameters as in Tab.~(\ref{tab:cfr}).
}
\label{tab:br_int}
\end{table}
